\documentclass[aps,pra,twocolumn,showpacs]{revtex4}

\input{epsf}

\begin{document}

\title{Quantum Monte Carlo study of quasi-one-dimensional Bose gases}

\author{G. E. Astrakharchik$^{(1,2)}$, D. Blume$^{(3)}$, S. Giorgini$^{(1)}$, and B. E. Granger$^{(4)}$}
\address{$^{(1)}$Dipartimento di Fisica, Universit\`a di Trento and BEC-INFM, I-38050 Povo, Italy\\
$^{(2)}$Institute of Spectroscopy, 142190 Troitsk, Moscow region, Russia\\
$^{(3)}$Department of Physics, Washington State University,
Pullman, Washington 99164-2814, USA\\
$^{(4)}$Department of Physics, Santa Clara University, Santa Clara, California
95053, USA}

\date{\today}

\begin{abstract}
We study the behavior of quasi-one-dimensional (quasi-1d) Bose gases
by Monte Carlo techniques, i.e., by the variational Monte Carlo,
the diffusion Monte Carlo, and the fixed-node diffusion Monte Carlo
technique. Our calculations confirm and extend our results of an earlier
study [Astrakharchik {\em{et al.}}, cond-mat/0308585]. 
We find that a quasi-1d Bose gas 
i) is well described by a 1d model Hamiltonian with
contact interactions and renormalized coupling constant; 
ii) reaches the Tonks-Girardeau regime
for a critical  value of the 3d scattering length $a_{3d}$; 
iii) enters a unitary regime for $|a_{3d}|\rightarrow \infty$, 
where the properties of the gas are 
independent of $a_{3d}$ and are similar to those 
of a 1d gas of hard-rods; and 
iv)  becomes 
unstable against cluster formation for a critical value of the
1d gas parameter. 
The accuracy
and implications of our results are discussed in detail.
\end{abstract}


\maketitle

\section{Introduction} 
\label{introduction}
Quasi-1d Bose gases have been realized in highly-elongated traps by tightly confining
the transverse motion of the atoms to their zero point oscillations \cite{EXP1D}. As compared 
to the 3d case, the role of quantum fluctuations is enhanced in 1d and these systems 
are predicted to exhibit peculiar properties, which 
cannot be described using traditional 
mean-field theories, but require more advanced many-body approaches. Particularly 
intriguing is the strong coupling regime, where, due to repulsion between particles, the
quasi-1d Bose gas behaves as if it consisted of fictitious 
spinless fermions.  This regime, the so called 
Tonks-Girardeau (TG) gas \cite{tonk36,Girardeau,Olshanii,Petrov},  has not been achieved 
yet, but is one of the main focus areas of present experimental investigations in this field 
\cite{EXP1D-1}. An interesting possibility to approach the strongly correlated TG regime is 
provided by magnetic field induced atom-atom Feshbach resonances \cite{EXPFR1,EXPFR2}. By utilizing this 
technique one can tune the 3d $s$-wave scattering length $a_{3d}$, and hence the strength 
of atom-atom interactions, to essentially any value, including zero and $\pm\infty$.     

Degenerate quantum gases near a Feshbach resonance have recently received a great deal
of interest both experimentally and theoretically. At resonance ($|a_{3d}|\to\infty$) 
the 3d scattering cross-section $\sigma$
is fixed by the unitary condition, $\sigma=4\pi/k^2$, 
where $k$ is the relative wave vector of the two atoms. In this regime it is predicted that, 
if the range $r_0$
of the atom-atom interaction potential is much smaller than the average interparticle 
distance, the behavior of the gas is universal, i.e., independent of the details of the 
interatomic potential and independent of the actual value of $a_{3d}$ \cite{Heiselberg,Cowell}. 
This is known as the unitary regime \cite{Ho}. In the case of 3d Bose gases, this unitary regime 
can most likely not be realized in experiments since three-body recombination is expected to set 
in when $a_{3d}$ becomes comparable to the average interparticle distance. Three-body recombination 
leads to cluster formation and hence makes the gas-like state unstable. The situation is different 
for Fermi gases, for which the unitary regime has already been reached experimentally \cite{EXPFR2}. 
In this case, the Fermi pressure stabilizes the system even for large $|a_{3d}|$. 

In quasi-1d 
geometries a new length scale becomes relevant, namely, the
oscillator length $a_\rho=\sqrt{\hbar/(m\omega_\rho)}$ of the tightly-confined transverse motion, where $m$ is the mass of the atoms and 
$\omega_\rho$ is the angular frequency of the harmonic trapping potential. For $|a_{3d}|\gg a_\rho$, the 
gas is expected to exhibit a universal behavior if the range $r_0$ 
of the atom-atom interaction potential is much 
smaller than $a_\rho$ and the mean interparticle distance is much larger than $a_\rho$. It has been 
predicted that three-body recombination processes are suppressed for strongly interacting 1d Bose 
gases \cite{Gangardt}. These studies raise the question whether the unitary regime can be reached 
in Bose gases confined in highly-elongated traps, that is, whether 
the quasi-1d bosonic gas-like
state is stable against cluster formation as $a_{3d}\to\pm\infty$.

This paper, which is an extension of an earlier study \cite{US}, 
investigates the properties 
of a quasi-1d Bose gas at zero temperature over a wide range of values of the 3d scattering length 
$a_{3d}$ using quantum Monte Carlo (MC)
techniques. We find that the system i) is well described by a 1d 
model Hamiltonian with contact interactions and renormalized coupling constant \cite{Olshanii} for 
any value of $a_{3d}$; ii) reaches the regime of a TG gas for a critical positive value of the 3d scattering length $a_{3d}$; 
iii) enters a unitary regime for large values of $|a_{3d}|$, 
that is, for 
$|a_{3d}| \rightarrow \infty$, where the properties of the quasi-1d 
Bose gas become independent of the actual value of $a_{3d}$ and are similar to those of a 1d gas of hard-rods; and iv) becomes unstable against cluster formation for a 
critical value of the 1d gas parameter, or equivalently,
for a 
critical negative value of the 3d scattering length $a_{3d}$.

The structure of this paper is as follows. Section~II discusses 
the energetics of two bosons in
quasi-1d harmonic traps. We review the mapping of the 3d Hamiltonian
to a 1d model Hamiltonian with contact
interactions and renormalized coupling constant~\cite{Olshanii}. 
The eigenenergies of the system
are calculated by exact diagonalization of both the 3d and the 
1d Hamiltonian. We use these results for two particles
to benchmark our quantum MC calculations presented in  Sec.~IV. Section~III 
discusses the relation between the 
3d and the 1d Hamiltonian for $N$ bosons under quasi-1d confinement. 
Section~IV discusses the quantum MC techniques 
used in the present study:
variational, diffusion and fixed-node diffusion MC. 
The trial wave functions used for the 
variational estimates and for importance sampling are discussed in detail. 
Section~V presents our
MC results for $N=2$ and $N=10$ atoms in 
highly-elongated harmonic traps over a wide 
range of values of the 3d scattering length $a_{3d}$. 
A comparison of the energetics of the lowest-lying 
gas-like state for the 3d and the 1d Hamiltonian is carried out.
In 
the $N=2$ case, we additionally compare with 
the essentially exact results presented in Sec.~II. 
In the $N=10$ case,
we additionally compare
with the energy of the lowest-lying gas-like state of the 1d Hamiltonian 
calculated using the local density
approximation (LDA). 
Section~VI  
discusses the stability of the
lowest-lying gas-like state against cluster formation 
when $a_{3d}$ is negative using the variational Monte Carlo (VMC) 
method. We provide a quantitative estimate of the criticality condition. 
Finally, Sec.~VII draws
our conclusions.

\section{Two Bosons under quasi-one-dimensional confinement}
\label{theoryn2}
Consider two interacting mass $m$ bosons with position vectors
$\vec{r}_1$ and $\vec{r}_2$, where $\vec{r}_i=(x_i,y_i,z_i)$, 
in a waveguide with harmonic confinement in the radial direction.
If we introduce the center of mass coordinate $\vec{R}$ and the relative 
coordinate $\vec{r}=\vec{r}_2-\vec{r}_1$, the problem separates.
Since the solution to the center of mass Hamiltonian is given
readily, we only consider the internal Hamiltonian $H_{3d}^{int}$, which can be
conveniently written in terms of cylindrical coordinates
$\vec{r}=(\rho,\phi,z)$,
\begin{eqnarray}
  H^{int}_{3d} = -\frac{\hbar^2}{2\mu} \nabla^2_{\vec{r}} + 
V(\vec{r}) + \frac{1}{2} \mu \omega_{\rho}^2 \rho^2, \label{eq_wgint3d}
\end{eqnarray}
where $\mu$ denotes the reduced two-body mass, $\mu=m/2$, and
$V(\vec{r})$ denotes the 
full 3d atom-atom interaction potential.

Considering
a regularized zero-range 
pseudo-potential $V(\vec{r}) = 2 \pi \hbar^2 a_{3d}/\mu \delta(\vec{r}) 
\frac{\partial}{\partial r} r$, where $a_{3d}$ denotes the
3d scattering length,
Olshanii~\cite{Olshanii} derives
an effective 1d Hamiltonian, 
\begin{eqnarray}
H^{int}_{1d}= - \frac{\hbar^2}{2 \mu} \frac{d^2}{d z^2}
+ g_{1d} \delta(z) + \hbar \omega_{\rho},\label{eq_wgint1d}
\end{eqnarray}
and renormalized coupling constant $g_{1d}$,
\begin{eqnarray}
g_{1d}=
\frac{2 \hbar^2 a_{3d}}{m a_\rho^2}
\left [ 1-|\zeta(1/2)|
\frac{a_{3d}}{\sqrt{2}a_\rho} \right]^{-1} \,,
\label{eq_g1dren}
\end{eqnarray}
which reproduce the low energy scattering solutions of the
full 3d Hamiltonian, Eq.~(\ref{eq_wgint3d}).
Here, 
$\zeta(\cdot)$ denotes the Riemann zeta function,
$\zeta(1/2) = -1.4604$.
Alternatively, $g_{1d}$ can be expressed through the effective 
1d scattering length $a_{1d}$~\cite{Olshanii}, 
\begin{eqnarray}
g_{1d}= -\frac{2\hbar^2}{ma_{1d}} ,\label{eq_g1da1d}
\end{eqnarray}
where 
\begin{eqnarray}
a_{1d} = - a_{\rho} \left( \frac{a_{\rho}}{a_{3d}} - 
\frac{|\zeta(1/2)|}{\sqrt{2}} \right) \,.
\label{eq_a1dren}
\end{eqnarray}
Olshanii's result shows 
that the waveguide gives rise to an effective interaction,
parameterized by the coupling constant $g_{1d}$,
which can be tuned to any strength by changing the ratio between
the 3d scattering length $a_{3d}$ and the transverse oscillator
length $a_{\rho}$.

Note that a recent K-matrix treatment~\cite{Granger} results in
an explicit energy dependence of the effective 1d coupling constant,
\begin{eqnarray}
g_{1d}(E^{int}_{3d})=
\frac{2 \hbar^2 a_{3d}(E^{int}_{3d})}{m a_\rho^2} \times \nonumber\\ 
\left[
1-\left|\zeta\left(\frac{1}{2},\frac{1}{2}\left(3-\frac{E^{int}_{3d}}{\hbar \omega_{\rho}}\right)\right)\right|
\frac{a_{3d}(E^{int}_{3d})}{ \sqrt{2} a_{\rho}} 
\right]^{-1} \,,
\label{eq_g1denergy}
\end{eqnarray} 
where $\zeta(\cdot,\cdot)$ denotes the Hurwitz zeta function and 
$E^{int}_{3d}$
the internal energy of the two boson system.
For scattering between two bosons with minimal internal energy, 
$E^{int}_{3d} = \hbar \omega_{\rho}$,
the Hurwitz zeta function in Eq.~(\ref{eq_g1denergy}) 
reduces to the Riemann zeta function,
since $\zeta(0.5,1)=\zeta(0.5)$.
With the additional assumption that the energy-dependence of
the 3d scattering length
$a_{3d}$ can be neglected, 
Eq.~(\ref{eq_g1denergy}) reduces to 
Eq.~(\ref{eq_g1dren}). For low-energy scattering the energy-independent 
effective 1d coupling constant, Eq.~(\ref{eq_g1dren}), is expected to provide 
a good description of the system, 
and we hence use it throughout this paper.

\begin{figure}[tbp]
\vspace*{-1.0in}
\centerline{\epsfxsize=3.25in\epsfbox{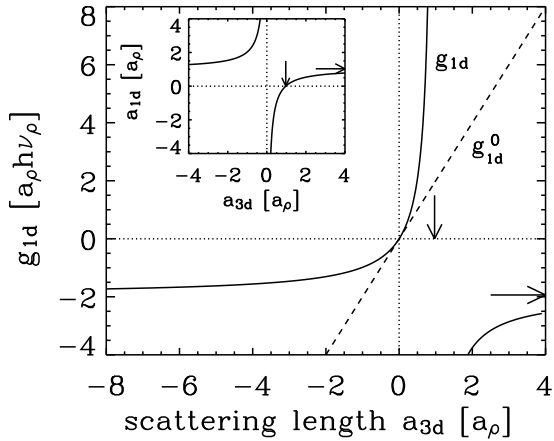}}
\vspace*{-.3in}
\caption{
One-dimensional coupling constants $g_{1d}$ 
[Eq.~(\protect\ref{eq_g1dren}),
solid line] and
$g_{1d}^{0}$ [Eq.~(\protect\ref{eq_g1dunren}), 
dashed line] 
as a function of the 3d scattering length $a_{3d}/a_{\rho}$.
The vertical arrow indicates the value of
$a_{3d}$ for which $g_{1d}$ diverges, $a_{3d}^{c}=0.9684a_{\rho}$.
The horizontal arrow indicates the asymptotic value of $g_{1d}$
as $|a_{3d}| \rightarrow \infty$,
$g_{1d}=-1.9368a_{\rho} \hbar \omega_{\rho}$. 
Inset: One-dimensional scattering length 
$a_{1d}$, Eq.~(\protect\ref{eq_a1dren}), 
as a function of $a_{3d}/a_{\rho}$.
The vertical arrow indicates the value of
$a_{3d}$ for which $a_{1d}$ goes through zero, $a_{3d}^{c}=0.9684a_{\rho}$.
The horizontal arrow indicates the asymptotic value of $a_{1d}$
as $|a_{3d}| \rightarrow \infty$,
$a_{1d}=1.0326a_{\rho}$.
The angular frequency $\omega_{\rho}$ determines the frequency $\nu_{\rho}$,
$\omega_{\rho}=2 \pi \, \nu_{\rho}$ (also, $\hbar \omega_{\rho}=h \nu_{\rho}$).
}
\label{fig1}
\end{figure}

The renormalized coupling constant, Eq.~(\ref{eq_g1dren}), 
can be compared with the
unrenormalized coupling constant $g_{1d}^0$,
\begin{eqnarray}
g_{1d}^0=\frac{2\hbar^2 a_{3d}}{m a_\rho^2}, \label{eq_g1dunren}
\end{eqnarray}
which is
obtained by
averaging the 3d coupling constant $g_{3d}=4\pi\hbar^2a_{3d}/m$ 
over the transverse oscillator ground 
state (see, e.g., Ref. \cite{Petrov}).
Figure~\ref{fig1} shows the unrenormalized coupling
constant $g_{1d}^0$ [dashed line, Eq.~(\ref{eq_g1dunren})] 
together with the renormalized coupling constant
[solid line, Eq.~(\ref{eq_g1dren})].
For $|a_{3d}| \ll a_{\rho}$, the 
renormalized coupling constant $g_{1d}$
is nearly identical to the
unrenormalized coupling constant $g_{1d}^0$.
For large $|a_{3d}|$, however, 
the confinement induced renormalization
becomes important, and the effective 1d coupling constant $g_{1d}$, 
Eq.~(\ref{eq_g1dren}), 
has to be used. 
At the critical value  
$a_{3d}^c=0.9684 a_{\rho}$ (indicated by
a 
vertical arrow in Fig.~\ref{fig1}),
$g_{1d}$ diverges.
For $a_{3d}\to\pm\infty$,
$g_{1d}$ 
reaches an asymptotic value, $g_{1d}=-1.9368a_{\rho} \hbar \omega_{\rho}$ 
(indicated by a
horizontal arrow in Fig.~\ref{fig1}).
Finally, $g_{1d}$ is negative for all negative 3d scattering lengths.
The inset of Fig.~\ref{fig1} shows the effective 1d scattering length $a_{1d}$,
Eq.~(\ref{eq_a1dren}), as a function of $a_{3d}$. For small positive
$a_{3d}$, $a_{1d}$ is negative and it changes sign at $a_{3d}=a_{3d}^c$
($a_{1d}=0$ for $a_{3d}=a_{3d}^c$).
Moreover, $a_{1d}$ reaches, just as $g_{1d}$, an asymptotic value for
$|a_{3d}| \rightarrow \infty$,
$a_{1d}=1.0326a_{\rho}$ (indicated by a horizontal
arrow in the inset of Fig.~\ref{fig1}). 
The renormalized 1d scattering length 
$a_{1d}$ is positive for negative $a_{3d}$, and approaches
$+\infty$ as 
$a_{3d}\rightarrow 0^-$. 
Figure~\ref{fig1} suggests that 
tuning the 3d scattering length $a_{3d}$ to large values
allows a universal quasi-1d regime, where $g_{1d}$ and $a_{1d}$
are independent of $a_{3d}$, to be entered.

The effective coupling constant $g_{1d}$, Eq.~(\ref{eq_g1dren}), 
has been derived 
for a wave guide geometry, that is, with no axial confinement.
However, it also describes
the scattering between two bosons
confined to other quasi-1d geometries.
Consider, e.g., a 
Bose gas under harmonic
confinement.
If the confinement in the axial direction
is weak compared to that of the radial direction, the
scattering properties of each atom pair
are expected to
be described accurately by the effective coupling constant 
$g_{1d}$ and the effective scattering 
length $a_{1d}$.

The internal motion of two bosons under highly-elongated
confinement can be described by the following 3d Hamiltonian
\begin{eqnarray}
  H^{int}_{3d} = -\frac{\hbar^2}{2 \mu} \nabla^2_{\vec{r}} +
V(\vec{r}) + 
\frac{1}{2} \mu \left(
\omega_{\rho}^2 \rho^2 + \omega_z^2\label{eq_h3dn2}
z^2 \right),
\end{eqnarray}
where $\omega_z$ denotes the angular frequency in the longitudinal direction,
$\omega_z= \lambda \omega_{\rho}$ ($\lambda$ denotes the aspect ratio,
$\lambda \ll 1$).  The eigenenergies $E_{3d}^{int}$ and eigenfunctions $\psi_{3d}^{int}$
of this Hamiltonian satisfy the Schr\"odinger equation, 
\begin{eqnarray}
H^{int}_{3d} \psi_{3d}^{int}(\rho,z)=
E^{int}_{3d} \psi_{3d}^{int}(\rho,z).\label{eq_se3dn2}
\end{eqnarray}
The corresponding 1d Hamiltonian reads
\begin{eqnarray}
H^{int}_{1d}= - \frac{\hbar^2}{2 \mu} \frac{d^2}{d z^2}
+ g_{1d} \delta(z) + \frac{1}{2} \mu \omega_z^2 z^2 + \hbar \omega_{\rho}.
\label{eq_h1dn2}
\end{eqnarray}
The 1d eigenenergies $E^{int}_{1d}$ 
of the time-independent Schr\"odinger
equation, 
\begin{eqnarray}
H^{int}_{1d} \psi_{1d}^{int}(z)=
E^{int}_{1d} \psi_{1d}^{int}(z),\label{eq_se1dn2}
\end{eqnarray}
can be determined semi-analytically
by solving the transcendental equation~\cite{Busch},
\begin{eqnarray}
g_{1d} = 2 \sqrt{2} 
\frac{\Gamma(\chi_z+1)}{\Gamma(\chi_z+1/2)} \; 
\tan(\pi \chi_z) \;
\hbar \omega_z \, a_z, \label{eq_trans}
\end{eqnarray}
self consistently for $\chi_z$ (for a given $g_{1d}$).
In the above equation, $\chi_z$ 
is an effective (possibly non-integer) quantum number, which determines 
the energy $E_z$,
\begin{eqnarray}
\chi_z=\frac{E_z}{2\hbar \omega_z}-\frac{1}{4}.\label{eq_nuz}
\end{eqnarray}
The energy $E_z$, in turn, determines the internal 1d eigenenergies
$E_{1d}^{int}$,
\begin{eqnarray}
E_{1d}^{int} = \lambda E_z + \hbar \omega_{\rho}.\label{eq_e1dint}
\end{eqnarray}
In Eq.~(\ref{eq_trans}), $a_z$ denotes the characteristic oscillator
length in the axial direction, $a_z = \sqrt{\hbar/(m \omega_z)}$.

\begin{figure}[tbp]
\vspace*{-1.0in}
\centerline{\epsfxsize=3.25in\epsfbox{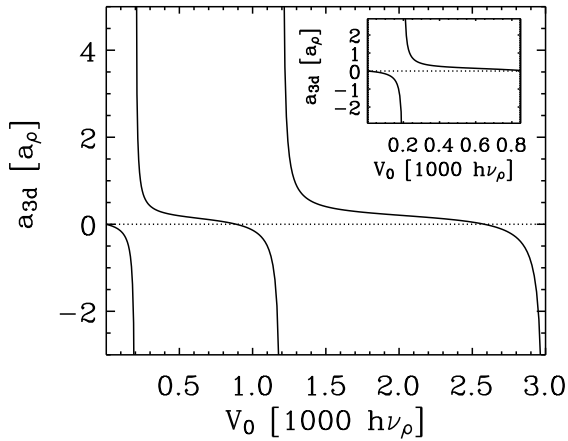}}
\vspace*{-.3in}
\caption{Three-dimensional 
$s$-wave scattering length $a_{3d}$ as a function of the 
well depth $V_0$ for the short-range model potential $V^{SR}$,
Eq.~(\protect\ref{eq_vsr}). Each time the 3d scattering length diverges
a new two-body $s$-wave bound state is created. 
Inset: Enlargement of the well depth region used in our calculations.}
\label{fig2}
\end{figure}

To compare the eigenenergies $E^{int}_{3d}$ and $E^{int}_{1d}$, 
we use, for the 3d atom-atom interaction potential $V(r)$,
a short-range (SR) model potential
$V^{SR}(r)$ 
that can support 
two-body 
bound states, 
\begin{eqnarray}
V^{SR}(r)=\frac{-V_0}{\mbox{cosh}^2(r/r_0)}.\label{eq_vsr}
\end{eqnarray}
In the above equation, $V_0$ denotes the well
depth, and
$r_0$ the range of the potential.
In our calculations, $r_0$ is fixed at a value much smaller than the 
transverse oscillator length, $r_0 = 0.1 a_\rho$.
To simulate the behavior of $a_{3d}$ near a field-dependent Feshbach resonance,
we vary the well depth $V_0$, and consequently, the
scattering length $a_{3d}$. 
It has been shown that such a
model describes many atom-atom scattering properties near a 
Feshbach resonance properly~\cite{tiesinga}.
Figure~\ref{fig2} shows the dependence of the 3d
scattering length $a_{3d}$ on $V_0$.
Importantly, $a_{3d}$
diverges
for particular values of the well depth $V_0$.
At each of these divergencies, 
a new two-body $s$-wave bound state is created. The inset of Fig.~\ref{fig2} shows
the range of well depths $V_0$ used in our calculations.

\begin{figure}[tbp]
\vspace*{-.2in}
\hspace*{0.15in}\centerline{\epsfxsize=4.75in\epsfbox{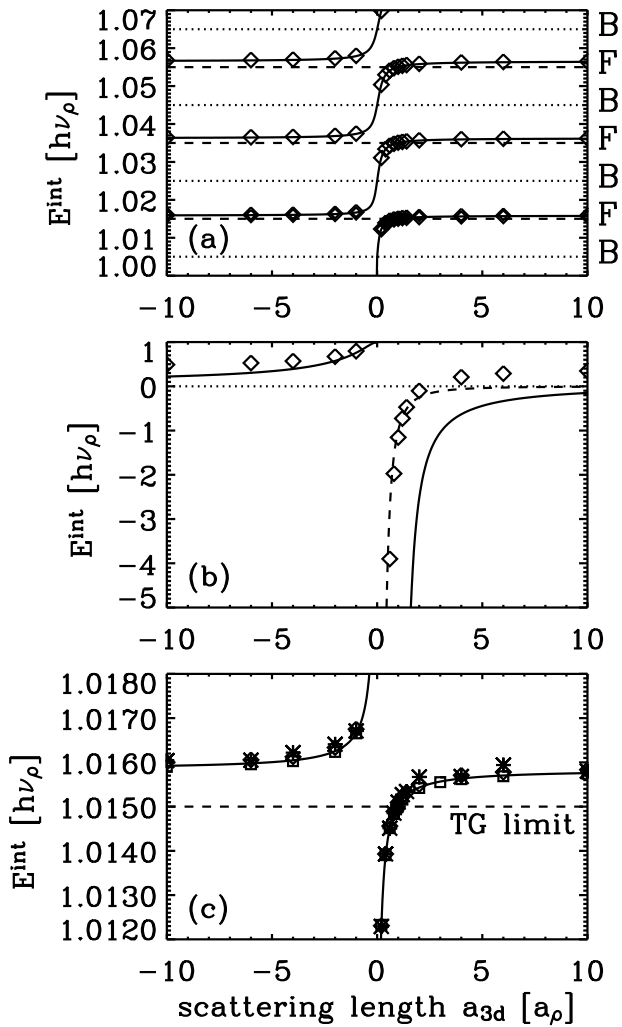}}
\caption{Internal eigenenergies $E^{int}$ 
as a function of the 3d scattering length $a_{3d}/a_{\rho}$
for two bosons under 
highly-elongated confinement with $\lambda=0.01$.
(a) 3d $s$-wave
eigenenergies $E^{int}_{3d}$ (diamonds) of gas-like states 
obtained
using the short-range model potential $V^{SR}$, 
Eq.~(\protect\ref{eq_vsr}), 
in a B-spline basis set calculation
together with
internal 1d eigenenergies 
$E^{int}_{1d}$ (solid lines).
Excellent agreement between the 3d and 1d energies is found.
Horizontal
dotted lines
show the lowest internal eigenenergies for two
non-interacting spin-polarized bosons,
while horizontal dashed lines
show those for two non-interacting spin-polarized fermions 
(indicated respectively by ``B'' and ``F'' on the right
hand side).
(b) 
Binding energy of molecular-like bound states.
In addition to the 3d and 1d energies [diamonds and solid lines,
respectively; see (a)],
a dashed line shows the 3d binding energy 
$-\hbar^2/(m a_{3d}^2)$.
(c) Enlargement of the lowest-lying gas-like state.
In addition to the 3d and 1d energies shown in (a),
asterisks show the 3d energies for the interaction
potential $V^{SR}$
calculated using the FN-DMC technique, and squares
the 1d energies for the contact interaction potential
calculated using the (FN-)\-DMC technique.
The statistical uncertainty of the (FN-)DMC 
energies is smaller than the symbol size.
Good agreement between the (FN-)DMC 
energies (asterisks and squares) and the energies
calculated by alternative
means (diamonds and solid lines) is found.
}
\label{fig3}
\end{figure}

To benchmark our MC calculations (see Secs.~\ref{montecarlo} 
and \ref{energetics}), 
we solve the 3d Schr\"odinger equation, Eq.~(\ref{eq_se3dn2}), 
with $\lambda=0.01$ for various
well depths $V_0$
using a two-dimensional B-spline basis in $\rho$ and $z$.
Figure~\ref{fig3} shows the resulting 3d eigenenergies $E^{int}_{3d}$ 
(diamonds) as a function of the 3d scattering length $a_{3d}$. 
We distinguish between two sets of states:
i) States with $E^{int}_{3d} \ge \hbar \omega_{\rho}$ are referred to as gas-like
states; their behavior is, to a good approximation, characterized
by the 3d scattering length $a_{3d}$, and is hence independent of
the detailed shape of the atom-atom potential. 
The energies of the gas-like states are
shown in Fig.~\ref{fig3}(a).
ii) States with $E^{int}_{3d}<\hbar \omega_{\rho}$ are referred to as
molecular-like
bound states; their behavior depends on the detailed shape of the atom-atom
potential. The energies of these bound states are
shown in Fig.~\ref{fig3}(b).
The well depth $V_0$ of the short-range 
interaction potential $V^{SR}$
is chosen such
that $V^{SR}$ supports ---
in the 
absence of the confining
potential --- no $s$-wave bound state for $a_{3d}<0$, and
one $s$-wave bound state for $a_{3d}>0$.
Figure~\ref{fig3}(b)
shows that the bound state remains bound for $|a_{3d} | \rightarrow \infty$
and for negative $a_{3d}$ if
tight radial confinement is present.
In addition, a dashed line shows the 3d binding energy, 
$-\hbar^2/(m a_{3d}^2)$,
which accurately describes the highest-lying molecular bound state
in the absence of any external confinement if $a_{3d}$ is much larger than the
range $r_0$ of the potential $V^{SR}$.

The B-spline basis calculations
yield not only the internal 3d eigenenergies $E_{3d}^{int}$,
but also the corresponding wave functions
$\psi_{3d}^{int}$. The
nodal surface of the lowest-lying gas-like state,
which is to
a good approximation an ellipse in the $\rho z$-plane,
is a crucial 
ingredient of our
many-body calculations. Sections~\ref{montecarlo} and 
\ref{energetics} discuss in detail how this nodal surface is used
to parametrize
our trial wave function entering the
MC calculations. 

To compare the energy spectrum for $N=2$ of the effective 1d Hamiltonian
with that
of the 3d Hamiltonian,
Fig.~\ref{fig3}
additionally shows the 1d eigenenergies $E_{1d}^{int}$
(solid lines) obtained by solving the Schr\"odinger equation
for $H_{1d}^{int}$, 
Eq.~(\ref{eq_h1dn2}),
semi-analytically [using the renormalized coupling constant
$g_{1d}$, Eq.~(\ref{eq_g1dren})].
Figure~\ref{fig3}(a) demonstrates excellent agreement 
between the 3d and the 1d internal energies 
for all states with gas-like character.
For positive $a_{3d}$, the effective 1d Hamiltonian fails 
to reproduce the energy spectrum 
of the molecular-like bound states of the 3d Hamiltonian
accurately [see Fig.~\ref{fig3}(b), and also Ref.~\cite{Bergeman,Bolda}].

Our main focus is in the lowest-lying energy
level with gas-like character. This energy branch is shown in
Fig.~\ref{fig3}(c) 
on an enlarged scale. A horizontal
dashed line
shows the lowest internal 3d eigenenergy 
for two non-interacting spin-polarized fermions (where the
anti-symmetry of the wave function enters in the $z$ coordinate).   
Our numerical calculations confirm~\cite{Bergeman} that 
for $a_{3d}=a_{3d}^c$ ($g_{1d} \rightarrow \infty$) 
the two boson system behaves as if it consisted of two 
non-interacting 
spin-polarized fermions (TG gas). The energy $E^{int}_{3d}$ 
is larger than that of 
two non-interacting fermions for $a_{3d}>a_{3d}^c$, and
approaches the first excited state energy of two non-interacting bosons
for $a_{3d} \rightarrow 0^-$ [indicated by a 
dotted line in Fig.~\ref{fig3}(a)].

For positive $g_{1d}$,
the 1d Schr\"odinger equation, Eq.~(\ref{eq_se1dn2}), 
does not support molecular-like bound states.
Consequently, the wave function of the lowest-lying gas-like state 
is positive definite everywhere. For negative $g_{1d}$, however,
one molecular-like two-body bound state exists. If $a_{1d}\ll a_z$ the bound-state 
wave function is
approximately
given by the eigenstate $\psi_{1d}^{int}$ of the 1d Hamiltonian without
confinement, Eq.~(\ref{eq_wgint1d}),
\begin{eqnarray}
\psi_{1d}^{int}(z)=\exp\left(-\frac{|z|}{a_{1d}}\right),
\end{eqnarray}
with eigenenergy $E_{1d}^{int}$,
\begin{eqnarray}
E_{1d}^{int}= -\frac{\hbar^2}{ma_{1d}^2} + \hbar \omega_{\rho}.
\label{eq_e1dbinding}
\end{eqnarray}
For the highly-elongated trap with $\lambda=0.01$ shown in Fig.~\ref{fig3}(b) 
and positive $a_{1d}$ the above 
binding energy nearly 
coincides with the exact eigenenergy of the molecular-like bound state
obtained from the solution of the transcendental equation~(\ref{eq_trans})
 (solid line). 
The two-body
binding energy, Eq.~(\ref{eq_e1dbinding}),
is largest for $a_{1d} \rightarrow 0^+$ ($g_{1d} \rightarrow -\infty$);
in this case,
the molecular-like bound state wave function is tightly-localized around
$z=0$, where $z=z_2-z_1$. Consider a system with $a_{1d} \ll a_z$.
For negative $g_{1d}$ (positive $a_{1d}$), the nodes along the relative
coordinate $z$ of the lowest-lying gas-like wave 
function (in this case, the first excited state) are then approximately 
given by $\pm a_{1d}$. 
Thus, imposing the
boundary condition $\psi_{1d}^{int}=0$ at $|z|= a_{1d}$ and 
restricting the configuration space to $z > a_{1d}$ allows
one to obtain an approximation to 
the eigenenergy of the first excited eigen state.
Furthermore, imposing the boundary condition $\psi_{1d}^{int}=0$ at $z=a_{1d}$ 
is identical to solving the 1d Schr\"odinger equation for a
hard-rod interaction potential
$V^{HR}(z)$,
\begin{eqnarray}
V^{HR}(z) = 
\left\{ \begin{array}{cll} \infty           & \mbox{ for } & z<a_{1d} \\
                              0            & \mbox{ for } & z \ge a_{1d}\;.
\end{array}\right.
\end{eqnarray}
For $N=2$, 
asterisks in Fig.~\ref{fig3}(c) show the fixed-node
diffusion Monte Carlo (FN-DMC) 
results obtained using the above fictitious hard-rod 
potential (see Sec.~\ref{energeticsn2}). 
Good agreement is found with the exact 1d 
eigenenergies obtained from Eqs.~(\ref{eq_trans})-(\ref{eq_nuz}). 
For $N>2$ bosons,
our 1d FN-DMC algorithm and our usage of the hard-rod equation of state
both take advantage
of a reduction of configuration space
similar to that discussed here for two bosons (see Secs.~\ref{theoryn} and
\ref{montecarlo}).

\section{N bosons under quasi-one-dimensional confinement}
\label{theoryn}
For tightly-confined trapped gases the 1d regime is reached if the transverse motion of the 
atoms is frozen, with all the particles occupying the ground state of the transverse harmonic
oscillator. At zero temperature, 
this condition requires that the energy per particle is dominated by the 
trapping potential, $E/N=\hbar\omega_\rho+\epsilon$, where 
the excess energy $\epsilon$ is much smaller 
than the separation between levels in the transverse direction, $\epsilon\ll \hbar\omega_\rho$.
In the following we consider situations where the Bose gas is in the 1d regime for any value 
of the 3d scattering length $a_{3d}$. For a fixed trap anisotropy parameter 
$\lambda$ and a fixed number of particles $N$ the above requirement is satisfied if $N\lambda\ll 1$. 
For $\lambda=0.01$ and $N=10$ (as considered in Sec.~\ref{energeticsn}) this condition is fulfilled.

To compare the 3d and 1d energetics of a Bose gas,
we consider the 3d and 1d 
Hamiltonian describing $N$ spin-polarized bosons,
\begin{eqnarray}
H_{3d}= \sum_{i=1}^N \left[ \frac{-\hbar^2}{2m}
\nabla_{\vec{r}_i}^2 + \frac{1}{2}m \left( \omega_{\rho}^2 \rho_i^2
+ \omega_z^2 z_i^2 \right) \right] + \nonumber \\
 \sum_{i<j}^N V(r_{ij}) \;, \label{eq_h3d}
\end{eqnarray}
and 
\begin{eqnarray}
H_{1d}=\sum_{i=1}^N \left( \frac{-\hbar^2}{2m}
\frac{\partial^2}{\partial z_i^2} + \frac{1}{2} m \omega_z^2 z_i^2 \right) 
+ g_{1d}\sum_{i<j}^N \delta(z_{ij}) + \nonumber \\ 
 N \hbar\omega_{\rho}  \; ,
\label{eq_h1d}
\end{eqnarray}
respectively.
The corresponding eigenenergies and eigenfunctions are given by solving
the Schr\"odinger equations,
\begin{eqnarray}
H_{3d} \psi_{3d}(\vec{r}_1,\cdots,\vec{r}_N)=
E_{3d} \psi_{3d}(\vec{r}_1,\cdots,\vec{r}_N) \label{eq_se3d}
\end{eqnarray}
and 
\begin{eqnarray}
H_{1d} \psi_{1d}(z_1,\cdots,z_N)=
E_{1d} \psi_{1d}(z_1,\cdots,z_N),\label{eq_se1d}
\end{eqnarray}
respectively.
In contrast to Sec.~\ref{theoryn2}, here we 
do not separate out the center of 
mass motion 
since the MC calculations used to solve the 3d and 1d
many-body Schr\"odinger equations 
can be most conveniently implemented in 
Cartesian coordinate space (see Sec.~\ref{montecarlo}).
In the following,
we refer to eigenstates of the confined 
Bose gas with energy greater than $N \hbar \omega_{\rho}$
as gas-like states, and to those with energy smaller than $N \hbar \omega_{\rho}$
as cluster-like bound states. 

Section~\ref{energeticsn} compares the energetics of the
lowest-lying gas-like
state of the 3d Schr\"odinger equation, Eq.~(\ref{eq_se3d}), obtained using 
the short-range potential V$^{SR}$, Eq.~(\ref{eq_vsr}),
with that obtained using the hard-sphere potential $V^{HS}$,
\begin{eqnarray}
V^{HS}(r) = 
\left\{ \begin{array}{cll} \infty           & \mbox{ for } & r<a_{3d} \\
                              0            & \mbox{ for } & r \ge a_{3d}\;.
\end{array}\right.
\end{eqnarray}
For $V^{HS}$, the $s$-wave
scattering length $a_{3d}$ coincides with the range of the potential.
For $V^{SR}$, in contrast, $r_0$ determines 
the range of the potential, while
the scattering length $a_{3d}$ is determined by $r_0$ and $V_0$.
For $a_{3d}\ll a_\rho$, both potentials 
give nearly identical results for the energetics
of the lowest-lying gas-like state, which depend to a 
good approximation
only on the value of 
$a_{3d}$. For $a_{3d}\gtrsim a_\rho$, instead, 
deviations due to the different
effective ranges become visible and only $V^{SR}$ 
yields results, which do not 
depend on the short-range details of the potential 
and which are compatible with a 1d 
contact potential. 

Section~\ref{energeticsn} also discusses the energetics of 
the 1d
Hamiltonian, Eq.~(\ref{eq_h1d}).
For small
$|g_{1d}|$, the energetics of the many-body 
1d Hamiltonian are described well by a 1d mean-field equation
with non-linearity. For negative $g_{1d}$, 
the mean-field framework describes, for example,  
bright solitons \cite{Carr,Kanamoto},
which
have been observed experimentally \cite{Hulet}.
For large $|g_{1d}|$, in contrast,
the system is highly-correlated, 
and any mean-field treatment will fail.
Instead, a many-body description that incorporates 
higher order correlations has to be used.
In particular, the limit $|g_{1d}| \rightarrow \infty$
corresponds to the strongly-interacting TG regime.

For infinitely strong particle interactions
($|g_{1d}| \rightarrow \infty$), Girardeau shows~\cite{Girardeau}, 
using the equivalence between the 1d $\delta$-function potential 
and a ``1d hard-point potential'',
that the
energy spectrum of the 1d Bose gas coincides with that of 
$N$ non-interacting spin-polarized fermions. 
The lowest eigenenergy per particle of the 1d Bose gas, Eq.~(\ref{eq_se1d}),
is, in the TG limit, given
by 
\begin{eqnarray}
\frac{E^{TG}_{1d}}{N}=\left(\frac{\lambda N}{2} + 1\right) \hbar \omega_\rho.
\label{eq_etg}
\end{eqnarray}
The corresponding gas density is given by the sum
of squares of single-particle wave functions,
\begin{eqnarray}
n^{TG}_{1d}(z)= \frac{1}{\sqrt{\pi} a_z} \sum_{k=0}^{N-1} \frac{1}{2^k k!}
H_k^2(z/a_z) \exp \left[ -(z/a_z)^2 \right] \; ,\label{eq_nf}
\end{eqnarray}
with the normalization $\int_{-\infty}^{\infty} n^{TG}_{1d}(z) dz = N$.
In Eq.~(\ref{eq_nf}), the $H_k$ denote Hermite polynomials,
and $z$ denotes the distance measured from the center of the
trap.
For large numbers of atoms, the density expression in Eq.~(\ref{eq_nf})
can be calculated using the LDA~\cite{Dunjko},
\begin{eqnarray}
n_{1d}^{TG}(z)= \frac{\sqrt{2N}}{ \pi a_{z}} 
\left(1-\frac{z^2}{2Na_z^2} \right)^{1/2} \, .\label{eq_ntg}
\end{eqnarray}
The above result
cannot reproduce
the oscillatory behavior of the exact density, Eq.~(\ref{eq_nf}),
but it does describe the overall behavior properly (see Sec.~\ref{stability}).

To characterize the {\em{inhomogeneous}} 1d Bose gas further, we 
consider
the many-body Hamiltonian
of the {\em{homogeneous}} 1d Bose gas,
\begin{eqnarray}
H_{1d}^{hom}=\sum_{i=1}^N \frac{-\hbar^2}{2m}
\frac{\partial^2}{\partial z_i^2} 
+ g_{1d}\sum_{i<j}^N \delta(z_{ij}) + N \hbar \omega_{\rho}\; .
\label{eq_h1dwg}
\end{eqnarray}
[By introducing the energy shift $N \hbar \omega_{\rho}$, 
our classification of gas-like states and
cluster-like bound states introduced 
after Eq.~(\ref{eq_se1d}) remains valid.]
For positive $g_{1d}$, $H_{1d}^{hom}$ corresponds to the Lieb-Liniger (LL)
Hamiltonian.
The gas-like states of the LL Hamiltonian, 
including its thermodynamic properties, have
been studied in detail \cite{Lieb}. 
The energy per particle of the lowest-lying gas-like state,
the ground state of the system, 
is given
by 
\begin{eqnarray}
\frac{E_{1d}^{LL}(n_{1d})}{N}= 
\frac{\hbar^2}{2m} e(\gamma) n_{1d}^2,\label{eq_ell}
\end{eqnarray}
where $n_{1d}$ denotes the density of the homogeneous system, and 
$e(\gamma)$ a function of the  
dimensionless parameter $\gamma= 2/(n_{1d} |a_{1d}|)$.

We use the known properties of the LL Hamiltonian to 
determine properties of the corresponding
inhomogeneous system, Eq.~(\ref{eq_h1d}),
within the LDA.
This approximation provides a correct description of the trapped gas
if the size of the atomic cloud is much larger than the characteristic 
length scale $a_z$ of the confinement in the longitudinal direction~\cite{Dunjko}.
Specifically, consider the local equilibrium condition,
\begin{eqnarray}
\mu(N)= \hbar \omega_{\rho} + \mu_{local}[n_{1d}(z)]+ \label{eq_mu}
\frac{1}{2}m \omega_z^2 z^2,
\end{eqnarray}
where $\mu_{local}(n_{1d})$ denotes the chemical potential of the 
homogeneous system with density $n_{1d}$,
\begin{eqnarray}
\mu_{local}(n_{1d}) = \frac{\partial\left[ 
n_{1d} E_{1d}^{LL}(n_{1d})/N
\right]}{\partial n_{1d}}. \label{eq_mulocal}
\end{eqnarray}
The chemical potential $\mu(N)$, Eq.~(\ref{eq_mu}),
can be calculated
using Eq.~(\ref{eq_mulocal}) together with the normalization
of the density, $\int_{-\infty}^\infty n_{1d}(z) dz = N$.
Integrating the chemical potential $\mu(N)$ then determines
the energy of the lowest-lying gas-like state
of the inhomogeneous $N$-particle system 
within the LDA.
The LDA treatment is
computationally less demanding than solving the many-body
Schr\"odinger equation, Eq.~(\ref{eq_se1d}), using MC techniques.
By comparing with our full 1d many-body results
we
establish the accuracy of the LDA (see Sec.~\ref{energeticsn}).

For negative $g_{1d}$, the Hamiltonian given in
Eq.~(\ref{eq_h1dwg}) supports cluster-like bound states
(``tightly-bound droplets''). 
The ground state energy and eigenfunction of the system 
are~\cite{McGuire}
\begin{eqnarray}
\frac{E_{1d}^{hom}}{N} = 
-\frac{\hbar^2}{6m a_{1d}^2} (N^2-1) +  \hbar \omega_{\rho},
\label{eq_e1dwg}
\end{eqnarray}
and 
\begin{eqnarray}
\psi_{1d}^{hom}(z_1,\cdots,z_N)= 
\prod _ {i<j}^N \exp \left(\frac{-|z_i-z_j|}{a_{1d}}  \right),
\label{eq_psi1dwg}
\end{eqnarray}
respectively.
The eigenstate given by Eq.~(\ref{eq_psi1dwg}) depends only on the relative
coordinates $z_{ij}$, that is, it is independent of the
center of mass of the system. 
Adding a confinement potential 
[see Eq.~(\ref{eq_h1d})] with $\omega_z$ such that $a_z \gg a_{1d}$ leaves 
the eigenenergy $E_{1d}^{hom}$ of this
cluster-like bound state to a good approximation
unchanged, while the corresponding wave function becomes 
localized at the center of the trap. This state
describes a bright soliton, whose energy can also
be determined within a mean-field framework~\cite{Kanamoto}.
An excited state of the many-body 1d Hamiltonian with confinement corresponds, 
e.g., to a 
state, where $N-1$ particles form a cluster-like
bound state, i.e., a soliton with $N-1$ particles, and where one particle
approximately occupies
the lowest harmonic oscillator state, i.e., has gas-like character. 
Similarly, molecular-like bound states can form with fewer particles. 

The above discussion implies that the lowest-lying gas-like state
of the 1d Hamiltonian with confinement, Eq.~(\ref{eq_h1d}) 
with negative $g_{1d}$,
corresponds to a highly-excited state.
For dilute 1d systems with negative $g_{1d}$, the
nodal surface of this excited state
can be
well approximated
by the following nodal surface: $\psi_{1d}=0$
for $z_{ij}=a_{1d}$, where $i,j=1,\cdots,N$ and $i<j$. 
As in the two-body case,
the many-body energy can then be 
calculated approximately by restricting the configuration space to
regions where the wave function is positive.
This corresponds to treating a gas of hard-rods of size $a_{1d}$.
In the low density limit,
we expect that the lowest-lying gas-like state of the 1d many-body
Hamiltonian with $g_{1d}<0$ is well described by a system of hard-rods
of size $a_{1d}$.

In addition to treating the full 1d many-body Hamiltonian,
we treat the inhomogeneous system with negative $g_{1d}$ 
within the LDA.
The equation of state of the {\em{uniform}} hard-rod gas with density
$n_{1d}$ is given by~\cite{Girardeau}
\begin{eqnarray}
\frac{E_{1d}^{HR}(n_{1d})}{N}=   \frac{\pi^2 \hbar^2 n_{1d}^2} 
{6m \, (1-n_{1d}a_{1d})^2}  +  \hbar \omega_{\rho} .
\label{eq_e1dhr}
\end{eqnarray}
We use this energy in the LDA treatment [see Eqs.~(\ref{eq_ell}) through 
(\ref{eq_mulocal})
with $E_{1d}^{LL}$ replaced by $E_{1d}^{HR}$].
The hard-rod equation of state treated within the LDA
provides a good description when $g_{1d}$ is negative,
but $|g_{1d}|$ not too small (see Secs.~\ref{energeticsn} and \ref{stability}).
To gain more insight, we
determine the expansion for inhomogeneous systems with $N \lambda \ll 1$
using the equation of state for the homogeneous hard-rod gas,
\begin{eqnarray}
\frac{E_{1d}}{N} - \hbar\omega_\rho = 
\hbar\omega_\rho \frac{N\lambda}{2}\left( 1 + \frac{128\sqrt{2}}{45\pi^2}
\sqrt{N\lambda}\frac{a_{1d}}{a_\rho} + \cdots \right) \;.
\label{eq_e1dexpansion}
\end{eqnarray}       
The first term corresponds to the energy per particle in the TG regime
[see Eq.~(\ref{eq_etg})];
the other terms can be considered as small corrections to the TG energy. 
In the unitary 
limit, that is, for $a_{1d}/a_\rho=1.0326$, expression (\ref{eq_e1dexpansion}) 
becomes independent of  
$a_{3d}$, and depends only on $N\lambda$.
Similarly, the linear density in the center 
of the cloud, $z=0$, is to lowest order given by the TG result, 
$n_{1d}^{TG}(0)=\sqrt{2N\lambda}/ (\pi a_\rho)$ [see Eq.~(\ref{eq_ntg})]. 
Section~\ref{stability} shows that the TG density provides a 
good description
of inhomogeneous 1d Bose gases over a fairly large range of negative $g_{1d}$.

\section{Quantum Monte Carlo methods}
\label{montecarlo}
This section describes 
the variational, diffusion and fixed-node 
diffusion MC methods 
(see, e.g.,~\cite{Guardiola}) used in the present study.  
These quantum MC techniques
solve the many-body Schr\"odinger equation for the ground state and 
for excited states at zero temperature.
Similar to other 
MC approaches, these techniques are based on stochastic numerical 
algorithms, which
are powerful when one is treating systems with many degrees of 
freedom. 

\subsection{Variational Monte Carlo method}
\label{vmc}
The VMC method was 
first introduced in the seminal work by
McMillan~\cite{McMillan} 
to study the ground state properties of liquid $^4$He. 
The VMC technique is 
based on the variational principle,  
\begin{equation}
\frac{\langle\psi_T|H|\psi_T\rangle}{\langle\psi_T|\psi_T\rangle} \ge E_0 \;,
\label{eq_ritz}
\end{equation}
where $\psi_T$ denotes a variational or trial wave function, 
which is parameterized in terms of a set of variational parameters.
In Eq.~(\ref{eq_ritz}), $H$ denotes the Hamiltonian of a bosonic $N$ particle 
system with ground state energy $E_0$ and stationary ground state
wave function $\Psi_0$.
The evaluation of the high-dimensional integral, Eq.~(\ref{eq_ritz}),
can be performed by MC techniques, resulting in the VMC energy 
expectation value.
This variational energy expectation value
is an upper bound to the exact ground state energy $E_0$.
Importantly, the variational principle also applies to excited states.
For a trial wave function $\psi_T$ with a given symmetry, 
the variational estimate provides an upper bound to the 
energy of the lowest 
excited state of the Hamiltonian $H$ with that symmetry. 

Choosing a good functional form for the trial wave function $\psi_T$
is a crucial step of the VMC method (and also of the DMC method
and the FN-DMC method,
see Secs.~\ref{dmc} and \ref{fndmc}). A general 
ansatz for $\psi_T$, which has been used successfully to describe 
a system of $N$ spinless bosons under external confinement
in either 3d or 1d~\cite{jastrow},
is a Bijl-Jastrow decomposition into
one- and two-body correlation factors. 
The one-body term accounts for the confining 
potential, while the two-body term accounts for interactions 
between particles. 

To describe the lowest-lying  gas-like state of $H_{3d}$ 
[Eq.~(\ref{eq_h3d})],
we use the following trial wave 
function 
\begin{equation}
\psi_T^{3d}(\vec{r}_1,...,\vec{r}_N)=\prod_{i=1}^{N}e^{-z_i^2/(2\alpha_z^2)}
e^{-\rho_i^2/(2\alpha_\rho^2)}\prod_{i<j}^N f_2(\vec{r}_i-\vec{r}_j) \,.
\label{VMC1}
\end{equation}    
Here, $\alpha_z$ and $\alpha_\rho$ determine 
the Gaussian width of $\psi_T^{3d}$ in the longitudinal and
transverse direction, respectively.
These variational parameters 
are optimized in the course of the VMC calculation by minimizing
the energy expectation value. The two-body correlation factor 
$f_2(\vec{r})$ is chosen to closely reproduce the scattering behavior
of two bosons at low energies.
For the hard-sphere potential, we use
\begin{equation}
f_2(\vec{r})=\left\{ \begin{array}{cll}          0           & \mbox{ for } &  |\vec{r}| \le a_{3d}   \\

                                       1-{a_{3d}}/{|\vec{r}|}   & \mbox{ for } &  |\vec{r}|>a_{3d} \;.
\end{array}\right.
\label{VMC2}
\end{equation}
The constraint $f_2=0$ for $r \le a_{3d}$ 
accounts for the 
boundary condition imposed by the hard-sphere potential; it is exact
even for the many-body system.
For the short-range potential, we use instead
\begin{equation}
f_2(\vec{r})=\left\{ \begin{array}{cll}          0 & \mbox{ for }          
&   \frac{\rho^2}{a^2}+\frac{z^2}{b^2} \le 1   \\
1-1/\sqrt{\frac{\rho^2}{a^2}+\frac{z^2}{b^2}}   & \mbox{ for } 
&  \frac{\rho^2}{a^2}+\frac{z^2}{b^2}>1 \;,
\end{array}\right.
\label{VMC3}
\end{equation}
where $a$ and $b$ denote the lengths of the semi-axes 
of an ellipse. 
For two particles under highly-elongated confinement,
the nodal surface is to a good approximation
ellipticly shaped (see Sec.~\ref{energeticsn2}).
Thus, the parameters $a$ and $b$ are determined by 
fitting the elliptical surface to the nodal surface
obtained by solving the Schr\"odinger equation for $N=2$, 
Eqs.~(\ref{eq_h3dn2}) and (\ref{eq_se3dn2}), by performing 
a B-spline basis set calculation. 
In contrast to $V^{HS}$, the constraint $f_2=0$ in Eq.~(\ref{VMC3})
parameterizes the many-body nodal surface for $V^{SR}$ 
only approximately. 
We expect that our parameterization leads to an accurate  description
of quasi-1d Bose gases
if the average distance between particles is much
larger than the semi-axes of the ellipse.
The trial wave functions discussed here in the context of our VMC
calculations also enter our (FN-)\-DMC calculations (see Sec.~\ref{dmc}
and \ref{fndmc}).

In the case of the 1d Hamiltonian,  Eq. (\ref{eq_h1d}),
we use a trial wave function of the form
\begin{equation}
\psi_T^{1d}(z_1,...,z_N)=\prod_{i=1}^N e^{-z_i^2/(2\alpha_z^2)}\prod_{i<j}^Nf_2(z_i-z_j) \;,
\label{VMC4}
\end{equation}      
where the Gaussian width $\alpha_z$ is 
treated as a variational parameter. The two-body correlation factor $f_2(z)$
is
chosen as 
\begin{equation}
f_2(z)=\left\{ \begin{array}{cll}  \cos[k_z(|z|-\bar{Z})]& \mbox{ for }             
&   |z| \le \bar{Z}   \\
1   &  \mbox{ for } & |z|>\bar{Z} \;.
\end{array}\right.
\label{VMC5}
\end{equation}
The cut-off length 
$\bar{Z}$ is fixed at $\bar{Z}=500 a_{1d}$, while
the wave vector $k_z$ is chosen such that
the boundary condition at $z=0$ imposed by the $\delta$-function 
potential
is satisfied:
$-k_z\tan(k_z \bar{Z})=1/a_{1d}$.
For negative $a_{1d}$ ($g_{1d}>0$) the correlation function, 
Eq.~(\ref{VMC5}), is
positive everywhere. 
For positive $a_{1d}$ ($g_{1d}<0$), in contrast, 
$f_2(z)$ changes sign at $|z|=a_{1d}$. The parameterization
given by Eq.~(\ref{VMC5}) is used in our 
stability analysis performed within a VMC framework (see Sec.~\ref{stability})
and in our DMC calculations
for $g_{1d}> 0$ (see 
Sec.~\ref{dmc}). 
To perform the FN-DMC calculations 
for
negative $g_{1d}$ (see Sec.~\ref{fndmc}),
we need to construct a trial wave function that is positive
definite everywhere.
In the FN-DMC calculations,
we thus use an alternative parameterization, which imposes the
constraint $f_2=0$ for $a_{1d}\le z$,
\begin{equation}
f_2(z)=\left\{ \begin{array}{cll}  0& \mbox{ for } &   z \le a_{1d}   \\
\cos[k_z(|z|-\bar{Z})]& \mbox{ for }             &  a_{1d} < z \le \bar{Z}   \\
                                       1   &  \mbox{ for } & z>\bar{Z} \;.
\end{array}\right.
\label{VMC6}
\end{equation}

\subsection{Diffusion Monte Carlo method}
\label{dmc}
The diffusion Monte Carlo (DMC) 
algorithm solves the time-independent 
Schr\"odinger equation of a $N$ particle system 
by introducing the imaginary time $\tau = it /\hbar$,
and propagating the function 
$f({\cal{R}},\tau)=\psi_T({\cal{R}})\Psi({\cal{R}},\tau)$ in imaginary time
$\tau$. 
Here, $\Psi({\cal{R}},\tau)$ denotes the wave function of the system,
which we are seeking.
The trial wave function $\psi_T({\cal{ R}})$, which can be optimized
in a series of VMC calculations,
enters the DMC calculations as input. 
In 3d, ${\cal{ R}}$ collectively denotes the position vectors,
${\cal{R}}=(\vec{r}_1,...,\vec{r}_N)$; in 1d, it collectively denotes the 
positions,
${\cal{R}}=(z_1,\cdots,z_N)$.

Using the function $f({\cal{R}},\tau)$, 
the time-dependent Schr\"odinger equation in imaginary time 
can be rewritten as
\begin{eqnarray}
-\frac{\partial f({\cal{R}},\tau)}{\partial \tau}= &-& D\nabla_{{\cal{R}}}^2 f({\cal{R}},\tau) 
+ D\nabla_{{\cal{R}}}
[F({\cal{R}}) f({\cal{R}},\tau)] \nonumber \\
&+& [E_L({\cal{R}})-E_{ref}] f({\cal{R}},\tau) \;,
\label{DMC1}
\end{eqnarray}
where
$E_L({\cal{ R}})=\psi_T({\cal{R}})^{-1}H\psi_T({\cal{R}})$ denotes 
the local energy, 
$F({\cal{R}})=2\psi_T({\cal{R}})^{-1}\nabla_{{\cal{R}}}\psi_T({\cal{R}})$ 
the quantum drift force,  and $D$ the diffusion constant, $D=\hbar^2/2m$.
The subscript ${\cal{R}}$ of the operator $\nabla$ indicates that 
the derivative has to be taken for every component of ${\cal{R}}$.
The constant energy shift $E_{ref}$ is introduced to
stabilize the
numerics. 
The solution of Eq.~(\ref{DMC1}) can be written formally as
\begin{equation}
f({\cal{R}}^\prime,\tau+\Delta \tau)=\int d{\cal{R}}\;G({\cal{R}}^\prime,{\cal{R}},\Delta \tau) 
f({\cal{R}},\tau) \;,
\label{DMC2}
\end{equation}
where 
the time-dependent Green's function $G$ is given by
\begin{equation}
G({\cal{R}}^\prime,{\cal{R}},\Delta \tau)=\langle{\cal{R}}^\prime|e^{-A \Delta \tau}|{\cal{R}}\rangle \;.
\end{equation}
Here, $A$ denotes the time evolution operator of Eq.~(\ref{DMC1}), 
$-\partial f({\cal{R}},\tau)/\partial \tau =
A f({\cal{R}},\tau)$.
If the short-time approximation to the
Green's function $G({\cal{R}}^\prime,{\cal{R}},\Delta \tau)$ is 
known (for sufficiently small $\Delta \tau$), 
the asymptotic solution for large times 
$\tau$, $f({\cal{R}},\tau \to\infty)$, can be obtained by
propagating $f$ for a large number of time steps $\Delta \tau$ (for more details see, e.g., \cite{Boronat}).
For a system of bosons, 
one can show that $f({\cal{R}},\tau\to\infty)=\psi_T({\cal{R}})\Psi_0({\cal{R}})$, where $\Psi_0$ is the exact 
ground-state eigenfunction. We calculate the eigenenergy $E_0$ 
by first propagating to large times $\tau$, and then 
averaging the local energy $E_L$ 
over the distribution $f({\cal{R}},\tau)$,
\begin{equation}
E_0=\frac{\int d{\cal{R}} \psi_T({\cal{R}}) H \Psi_0({\cal{R}})}{\int d{\cal{R}} \psi_T({\cal{R}}) 
\Psi_0({\cal{R}})}
=\frac{\int d{\cal{R}} f({\cal{R}},\tau \to\infty) E_L({\cal{R}})}{\int d{\cal{R}} 
f({\cal{R}},\tau \to\infty)} \;.
\end{equation} 
Apart from statistical errors, the DMC method determines the 
energy of the nodeless ground state of a 
system of $N$ bosons essentially exactly. 
Importantly, the energy expectation value calculated by the DMC
technique with importance sampling, which we use in our
study, is independent of 
the detailed shape of $\psi_T$ as long as $\psi_T$ is positive definite
everywhere.

We use the outlined algorithm to 
calculate the ground-state energy $E_{3d}$ of the 3d Hamiltonian, 
Eq.~(\ref{eq_h3d}), for the hard-sphere interaction potential, $V^{HS}$, 
and that of the 1d Hamiltonian, Eq.~(\ref{eq_h1d}), 
for positive coupling constant, $g_{1d}>0$. 
The trial wave function $\psi_T$
used in these 
cases is given by Eqs.~(\ref{VMC1}) and (\ref{VMC2}), 
and by Eqs.~(\ref{VMC4}) and (\ref{VMC5}), respectively. 

The DMC algorithm cannot be used directly to calculate excited states.
If $\Psi$ is an excited state, that is, 
if $\Psi$ is orthogonal to $\Psi_0$, 
the function 
$f({\cal{R}},\tau)=\psi_T({\cal{R}})\Psi({\cal{R}},\tau)$ 
is not positive everywhere in configuration
space and can hence not be interpreted as a 
probability density. This leads to the
so-called sign problem. 
To nevertheless calculate excited state energies
such as
the energy of 
the lowest-lying gas-like state of the 3d Hamiltonian, Eq.~(\ref{eq_h3d}), 
with the short-range interatomic potential $V^{SR}$, Eq.~(\ref{eq_vsr}), 
or of the 1d Hamiltonian, Eq.~(\ref{eq_h1d}) with $g_{1d}<0$,
we apply
the FN-DMC method.

\subsection{Fixed-node diffusion Monte Carlo method}
\label{fndmc}
The 
FN-DMC method~\cite{Reynolds} 
modifies the DMC method to allow approximate treatment of excited states
of many-body systems.
The idea of the FN-DMC method
is to treat excited states
by ``enforcing'' the positive definiteness of 
the probability 
distribution $f({\cal{R}},\tau)=\psi_T({\cal{R}})\Psi({\cal{R}},\tau)$.
The function $f$ is positive definite 
everywhere in configuration space, and can hence be interpreted
as a probability distribution,
if $\psi_T$ and $\Psi$ 
change sign together, and thus share the same (high-dimensional) nodal
surface.
To ensure positive definiteness of $f$, the
trial wave function $\psi_T$ imposes a nodal constraint, which is fixed
during the calculation. 
Within this constraint, the function $f$ is propagated (following 
a scheme very similar to that outlined in Sec.~\ref{dmc}), and reaches 
an asymptotic distribution for large $\tau$,
$f({\cal{R}},\tau \to\infty)=
\psi_T({\cal{R}})\Psi({\cal{R}})$.
In the FN-DMC method, 
$\Psi$ is an approximation to the exact excited 
eigenfunction 
of the many-body Schr\"odinger equation (and not the exact eigenfunction as
in the DMC method). 
It can be proven that, due to the nodal constraint, the fixed-node 
energy is a variational upper bound to the 
exact eigenenergy for a given symmetry~\cite{Reynolds}.
In particular, if the nodal surface of $\psi_T$ were exact, 
then $\Psi$ would be exact. Thus, the FN-DMC energy 
depends crucially on a good parameterization of the many-body nodal surface.

Section~\ref{energetics} reports the energy of the lowest-lying gas-like state 
of the 3d Hamiltonian, Eq.~(\ref{eq_h3d}), for the short-range
potential $V^{SR}$, and that of the 1d Hamiltonian, Eq.~(\ref{eq_h1d}), 
for $g_{1d}<0$ calculated using the FN-DMC method. 
As discussed in Sec.~\ref{vmc},
the nodal surface of the many-body trial wave function $\psi_T$ 
is constructed from the essentially exact nodal surface of 
the two-body wave function describing the lowest-lying gas-like state. 
In the 3d case, the importance sampling 
trial wave function is given by Eqs.~(\ref{VMC1}) and (\ref{VMC3}). 
In the 1d case,
the importance sampling 
trial wave function is instead given by Eqs.~(\ref{VMC4}) and (\ref{VMC6}). 
We expect that the FN-DMC 
approach implemented here
results in accurate many-body energies
of dilute quasi-1d Bose gases.

\section{Energetics of quasi-one-dimensional Bose gases}
\label{energetics}
Table~\ref{tab1} summarizes the techniques 
used
to solve the 1d and 3d Hamiltonian, respectively.
This table is meant to guide the reader through our result
sections.
Section~\ref{energeticsn2} discusses
our MC energies for two-particle systems, while 
Sec.~\ref{energeticsn} discusses the energetics for larger systems,
calculated within various frameworks.
Finally, Sec.~\ref{stability} discusses the stability of quasi-1d Bose gases.

\begin{table}
\begin{tabular}{l|l|l|l} 
Hamiltonian & interaction & technique & Section \\ \hline
$H_{3d}$ & $V^{HS}$ & DMC 	& VB \\
$H_{3d}$ & $V^{SR}$ & FN-DMC 	& VA, VB \\ \hline
$H_{1d}$ & $g_{1d}>0$ & DMC 	& VA, VB \\
$H_{1d}$ & $g_{1d}>0$ & LDA, LL & VB  \\
$H_{1d}$ & $g_{1d}<0$ & FN-DMC 	& VA, VB \\
$H_{1d}$ & $g_{1d}>0$ & LDA, hard-rod & VB \\
$H_{1d}$ & $g_{1d}<0$ & VMC 	& VI \\
\end{tabular}
\caption{
Guide that summarizes the techniques used to solve
the 3d and 1d Hamiltonian, Eqs.~(\protect\ref{eq_h3d}) and 
(\protect\ref{eq_h1d}), 
respectively. Column 2 specifies the
atom-atom
interactions of the many-body Hamiltonian,
column 3 lists the techniques used to solve
the corresponding many-body Schr\"odinger equation, and column
4 lists the sections that discuss the results obtained
using this approach.}
\label{tab1}
\end{table}

\subsection{Two-body system}
\label{energeticsn2}
Section~\ref{theoryn2} discusses the calculation of
the energy spectrum related to the internal 
motion of two bosons under highly-elongated confinement, Eq.~(\ref{eq_h3dn2}), 
using a B-spline basis, and the eigenspectrum  related to the internal motion
of two bosons
under 1d confinement, Eq.~(\ref{eq_h1dn2}), 
using Eqs.~(\ref{eq_trans}) through (\ref{eq_e1dint}).
We now use these essentially exact 
eigenenergies to benchmark our (FN-)DMC calculations.
Toward this end, we solve the
3d Schr\"odinger equation, Eq.~(\ref{eq_h3d}), and the 1d Schr\"odinger
equation, Eq.~(\ref{eq_h1d}), for various interaction strengths
for $N=2$ and $\lambda=0.01$ using (FN-)DMC techniques.
The resulting MC energies $E_{3d}$ and $E_{1d}$ include the center of mass
energy of $(1 + \lambda/2)\hbar \omega_{\rho}$.
To compare with the internal eigenenergies discussed in Sec.~\ref{theoryn2},
we subtract the center of mass energy from the (FN-)DMC energies.

\begin{figure}[tbp]
\vspace*{-1.0in}
\centerline{\epsfxsize=3.25in\epsfbox{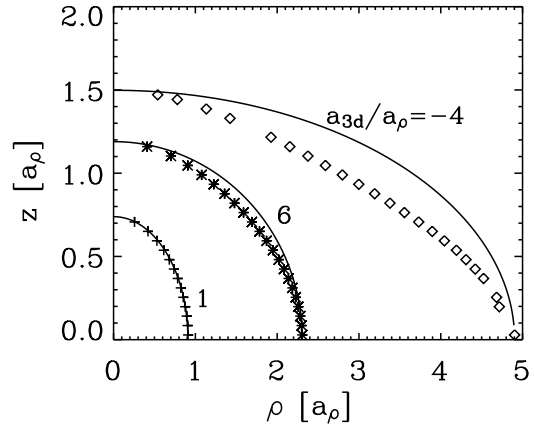}}
\vspace*{-.3in}
\caption{
Nodal surface
of the 
trial wave function $\psi_T$ [solid lines, Eq.~(\protect\ref{VMC3})]
together with the essentially exact nodal surface calculated 
using a B-spline basis set 
(symbols) 
for $\lambda=0.01$, $N=2$,
and three different
scattering lengths, 
$a_{3d}/a_{\rho}=1$ (pluses),
$a_{3d}/a_{\rho}=6$ (asterisks),
and 
$a_{3d}/a_{\rho}=-4$ (diamonds).
The nodal surface is shown as a function of the internal
coordinates $z$ and $\rho$.
Good agreement 
between the elliptical nodal surface (solid lines)
and the essentially exact nodal surface (symbols)
is visible for $a_{3d}/a_{\rho}=1$ and $6$. Small deviations 
are visible for $a_{3d}/a_{\rho}=-4$.
}
\label{fig5}
\end{figure}

For $N=2$, the lowest-lying gas-like state of the 
3d Hamiltonian for the short-range potential $V^{SR}$
is the first excited eigenstate.
Consequently, we solve the 3d Schr\"odinger equation by the FN-DMC technique
using the trial wave function $\psi_T$ given by Eqs.~(\ref{VMC1})
and (\ref{VMC3}).
Figure~\ref{fig5} shows the elliptical nodal surface of the 
trial wave function $\psi_T$ (solid lines)
together with the essentially exact nodal surface calculated 
using a B-spline basis set 
(symbols; see also Sec.~\ref{theoryn2}) for $\lambda=0.01$
and three different
scattering lengths, $a_{3d}/a_{\rho}=1,6$, and $-4$.
Notably, the semi-axes $a$ along the $\rho$ coordinate is larger than that 
along the $z$ coordinate ($a/b >1$),
``opposing'' the shape of the confining potential,
for which the characteristic length along the $\rho$ coordinate
is smaller than that along the $z$ coordinate ($a_{\rho}/a_z<1$).
Figure~\ref{fig5} indicates good agreement 
between the essentially exact nodal surface
and the parameterization of the nodal surface by an ellipse 
for  $a_{3d}/a_{\rho}=1$ and $6$;
some discrepancies become apparent for negative $a_{3d}$.
Since the FN-DMC method results in the exact energy if the nodal
surface of $\psi_T$ coincides with the nodal surface of the exact 
eigenfunction, comparing the FN-DMC energies for two particles
with those obtained from a B-spline basis set calculation provides a direct
measure of the quality of the nodal surface of $\psi_T$.
Figure~\ref{fig3}(c) compares the internal 3d energy of the
lowest-lying gas-like state calculated using a B-spline basis
(diamonds, see Sec.~\ref{theoryn2}) 
with that calculated using the FN-DMC technique (asterisks).
The agreement between these two sets of energies
is --- within the statistical uncertainty ---
excellent for all scattering lengths $a_{3d}$ considered.
We conclude that our 
parameterization of the two-body nodal surface, Eq.~(\ref{VMC3}), 
is accurate over
the whole range of interaction strengths $a_{3d}$ considered.

\begin{figure}[tbp]
\vspace*{-1.0in}
\centerline{\epsfxsize=3.25in\epsfbox{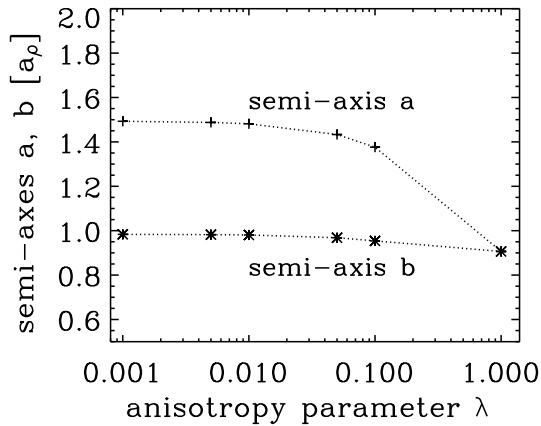}}
\vspace*{-.3in}
\caption{
Semi-axes $a$ (pluses) and $b$ (asterisks) obtained by fitting
an ellipse [see Eq.~(\protect\ref{VMC3})] to the
essentially exact nodal surface for two bosons under cylindrical
confinement,
calculated 
using a B-spline basis set as a function
of the anisotropy parameter $\lambda$,  
for $a_{3d}/a_{\rho}=2$. 
Dotted lines are shown to guide the eye.
For $\lambda \le 0.01$, the nodal surface is nearly independent of
the anisotropy parameter $\lambda$.
}
\label{fig6}
\end{figure}

We expect that our parameterization of the two-body nodal
surface is to a good approximation independent of the confining potential
in $z$ (for small enough $\lambda$). In fact, we expect our
nodal surface to closely resemble that of the scattering wave function 
at low scattering energy of 
the 3d wave guide Hamiltonian given by Eq.~(\ref{eq_wgint3d}).
To quantify this statement, Fig.~\ref{fig6}
shows the semi-axes $a$ and $b$ (pluses and asterisks,
respectively) obtained by fitting
an ellipse, see Eq.~(\ref{VMC3}), to the nodal
surface obtained by solving the Schr\"odinger equation for
the two-body Hamiltonian, Eq.~(\ref{eq_h3dn2}), 
using a B-spline basis 
for various
aspect ratios $\lambda$ ($\lambda=0.001,\cdots,1$), and
fixed scattering length, $a_{3d}=2a_{\rho}$
(similar results are found for other scattering lengths).
Indeed, the nodal surface for a given $a_{3d}/a_{\rho}$ 
is nearly independent of
the aspect ratio $\lambda$ for $\lambda \le 0.01$.
These findings for two particles imply that the parameterization of the 
nodal surface of $\psi_T$ used in the FN-DMC many-body calculations
should be good as long as the density along $z$ is small.

Next, consider the 1d Hamiltonian, Eq.~(\ref{eq_h1d}), for $N=2$.
For positive
$g_{1d}$, the lowest-lying gas-like state is the ground state of the 
two-body system and we hence use the DMC technique [with $\psi_T$
given by Eqs.~(\ref{VMC4}) and (\ref{VMC5})]; 
for $g_{1d}<0$, the lowest-lying gas-like state is the first
excited state, and we instead use the FN-DMC
technique [with $\psi_T$
given by Eqs.~(\ref{VMC4}) and (\ref{VMC6})]. 
Figure~\ref{fig3} shows the 1d energies of the lowest-lying 
gas-like state calculated 
using Eqs.~(\ref{eq_trans}) through (\ref{eq_e1dint}) 
(solid line), together with those
calculated by the (FN-)\-DMC technique (squares). 
We find excellent agreement between these two sets of 1d energies.

The comparison for two bosons between  
the (FN-)\-DMC energies and the energies calculated by
alternative means serves as a stringent test of our MC codes,
since these codes are implemented such that the number of particles
enters simply as a parameter. 

\subsection{N-body system}
\label{energeticsn}
This section presents our many-body study, which investigates
the properties of quasi-1d Bose gases over a wide range of 
scattering lengths $a_{3d}$. We focus specifically on three 
distinct regimes:
i) $0<a_{3d}<a_{3d}^c$ ($g_{1d}$ is positive);
ii) $|a_{3d}| \rightarrow \infty$ ($g_{1d}$ and $a_{1d}$ are independent
of $a_{3d}$; unitary regime); and
iii) $a_{3d} \rightarrow 0^-$ ($a_{1d}$ is large and positive;
onset of instability).
We discuss the energetics of quasi-1d
Bose gases for $N=10$. Our results presented here
support
our earlier conclusions, which are based on a study conducted 
for a smaller system,
i.e., for $N=5$~\cite{US}.

For small $\lambda$ (here, $\lambda=0.01$), the radial 
angular frequency $\omega_{\rho}$ dominates
the eigenenergies of the 3d and of the 1d Schr\"odinger equation.
The shift of the eigenenergy
of the lowest-lying gas-like state as a 
function of the interaction strength is,
however, set by the axial angular frequency $\omega_z$.
To emphasize the dependence of the eigenenergies of the
lowest-lying gas-like state on $\omega_z$,
we report 
the energy per particle subtracting the constant offset
$\hbar \omega_{\rho}$, that is, we report the 
quantity $E/N-\hbar \omega_{\rho}$.

\begin{figure}[tbp]
\vspace*{-.2in}
\centerline{\epsfxsize=3.25in\epsfbox{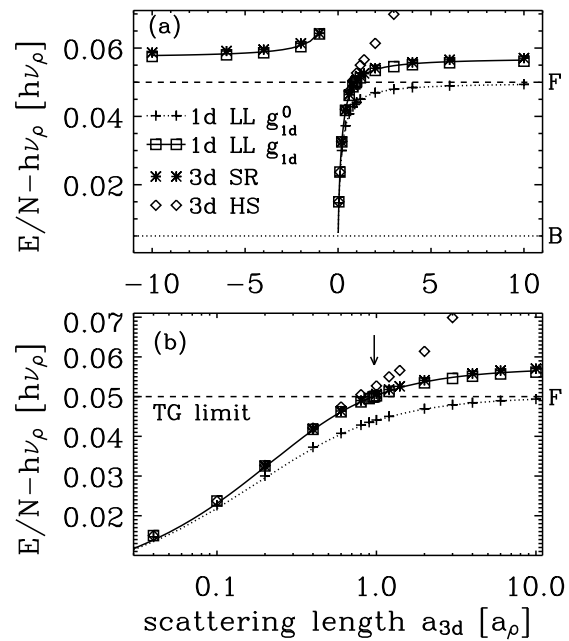}}
\vspace*{-.3in}
\caption{Three-dimensional 
(FN-)DMC energy per particle, $E_{3d}/N-\hbar \omega_{\rho}$,
calculated using
$V^{HS}$ (diamonds) and $V^{SR}$ (asterisks), respectively,
together with 1d (FN-)DMC energy per particle, 
$E_{1d}/N-\hbar \omega_{\rho}$, calculated using
$g_{1d}$ [squares, Eq.~(\protect\ref{eq_g1dren})] and 
$g_{1d}^{0}$ [pluses, Eq.~(\protect\ref{eq_g1dunren})],
respectively, as a function of $a_{3d}$ [(a) linear scale;
(b) logarithmic
scale] for $N=10$ and $\lambda=0.01$.
The statistical uncertainty of the (FN-)DMC energies is smaller than
the symbol size. 
Dotted and solid lines show the 1d energy per particle calculated
within the LDA for $g_{1d}^0$, 
Eq.~(\protect\ref{eq_g1dunren}) [using the LL equation of
state] and for $g_{1d}$, Eq.~(\protect\ref{eq_g1dren})
[using the LL equation of
state for $g_{1d}>0$, and the hard-rod equation of state for
$g_{1d}<0$], respectively. 
A dotted horizontal line indicates the energy per particle
of a non-interacting Bose gas, and
a dashed horizontal line indicates the
TG energy per particle. A vertical arrow the position where $g_{1d}$, 
Eq.~(\protect\ref{eq_g1dren}),
diverges.} \label{fig4}
\end{figure}

Consider the lowest-lying gas-like state of the
3d Schr\"odinger equation. 
Figure~\ref{fig4} shows the 3d energy per particle, 
$E_{3d}/N-\hbar \omega_{\rho}$, as a function of $a_{3d}$ for $N=10$ 
under 
quasi-1d confinement, 
$\lambda=0.01$, for the hard-sphere two-body potential $V^{HS}$ (diamonds) 
and the short-range 
potential
$V^{SR}$ (asterisks).
The energies for $V^{HS}$
are calculated using the DMC method 
[with $\psi_T$ given
by Eqs.~(\ref{VMC1}) and (\ref{VMC2})], 
while those for 
$V^{SR}$ are calculated using the 
FN-DMC method [with $\psi_T$ given
by Eqs.~(\ref{VMC1}) and (\ref{VMC3})].
For small $a_{3d}/a_{\rho}$, the energies for these 
two two-body potentials agree within the statistical uncertainty. For $a_{3d}\gtrsim a_{\rho}$, however, clear 
discrepancies are visible. The DMC energies for $V^{SR}$ cross the TG energy per particle 
(indicated by a dashed horizontal line), 
$E/N-\hbar\omega_\rho=\hbar\omega_{\rho} \lambda N/2$, very close to the value $a_{3d}^{c}=0.9684 a_{\rho}$ 
[indicated by a vertical arrow in Fig.~\ref{fig4}(b)], 
while the energies for $V^{HS}$ cross the TG energy per 
particle at a somewhat smaller value of
$a_{3d}$. 
For $V^{HS}$, ``fermionization'' of a quasi-1d gas has
previously been investigated in detail~\cite{dmcfermi}. The present paper
goes beyond these previous studies in that we consider a short-range 
potential, whose applicability extends to the regime $a_{3d} > a_{3d}^c$.

For
$a_{3d}>a_{3d}^c$, the energy 
for the short-range potential $V^{SR}$ of the lowest-lying gas-like state
increases  slowly with increasing $a_{3d}$, and becomes
approximately constant for large values of $|a_{3d}|$. The
limit $|a_{3d}| \rightarrow \infty$ corresponds to 
the unitary regime (see below). Notably,
the 3d energy behaves smoothly as $a_{3d}$ diverges.
The 3d energy slowly increases further
for increasing negative $a_{3d}$, and changes more rapidly
as $a_{3d} \rightarrow  0^-$. The $|a_{3d}| \rightarrow \infty$
regime and the $a_{3d} \rightarrow 0^-$ regime are discussed in more detail
below.

To compare our results obtained for the 3d Hamiltonian, 
$H_{3d}$, with those for the 1d Hamiltonian, $H_{1d}$, 
we also solve the Schr\"odinger equation for $H_{1d}$, Eq.~(\ref{eq_h1d}),
for the lowest-lying gas-like state.
For positive coupling constants, 
$g_{1d}>0$, the lowest-lying gas-like state is the many-body ground state, and
we hence use the DMC method [with $\psi_T$ given
by Eqs.~(\ref{VMC4}) and (\ref{VMC5})].
For $g_{1d}<0$, however, the 1d 
Hamiltonian supports cluster-like bound states. In this case,
the lowest-lying gas-like state corresponds to an excited
many-body state, and we hence solve the 1d Schr\"odinger equation 
by the FN-DMC method [with $\psi_T$ given
by Eqs.~(\ref{VMC4}) and (\ref{VMC6})].

Figure~\ref{fig4} shows 
the resulting 1d energies per particle, $E_{1d}/N-\hbar \omega_{\rho}$,
for the renormalized 
coupling constant $g_{1d}$ [squares, 
Eq.~(\ref{eq_g1dren})], and the unrenormalized 
coupling constant $g_{1d}^0$ [pluses, Eq.~(\ref{eq_g1dunren})], respectively. 
The 1d energies calculated using the two 
different coupling constants agree well for small $a_{3d}$, 
while clear discrepancies become apparent for $a_{3d} \gtrsim a_{3d}^c$.
In fact, the 1d energies calculated using the unrenormalized coupling constant
$g_{1d}^0$ approach the TG energy (dashed horizontal line) asymptotically 
for $a_{3d}\rightarrow \infty$, but do not become larger than 
the TG energy.
The 1d energies calculated using the 
renormalized 1d coupling constant $g_{1d}$ agree 
well with the 3d energies calculated using  
the short-range 
potential $V^{SR}$ (asterisks) up to very large values of the 3d 
scattering length $a_{3d}$.
In contrast, 
the 
1d energies deviate clearly from the 3d energies calculated using  
the hard-sphere
potential $V^{HS}$ (diamonds) at large $a_{3d}$.

The 1d energies calculated using the renormalized coupling constant 
agree with the 3d energies calculated using the short-range potential
$V^{SR}$ also for $|a_{3d}| \rightarrow \infty$, that is,
in the unitary regime, 
and for negative $a_{3d}$. 
Small deviations between the 1d energies calculated using the renormalized 
1d coupling constant $g_{1d}$ and the 3d energies calculated using  
the short-range 
potential $V^{SR}$ are visible; 
we attribute these to the finite range of $V^{SR}$. 
The deviations should decrease with decreasing range
$r_0$ of the short-range potential $V^{SR}$. On the other hand,
$r_0$ determines to first order
the energy-dependence of the scattering length $a_{3d}$. Thus,
usage of an energy-dependent coupling constant $g_{1d}$ 
[see Eq.~(\ref{eq_g1denergy})]
should also reduce the deviations between the 1d 
energies 
and the 3d energies calculated using  
the short-range 
potential $V^{SR}$~\cite{Granger}. 
Such an approach is, however, beyond the scope of this
paper.

We conclude that the renormalization of the effective 1d 
coupling constant $g_{1d}$ and of the 1d scattering length $a_{1d}$ 
are crucial to reproduce 
the results of the 3d Hamiltonian $H_{3d}$ when 
$a_{3d}\gtrsim a_\rho$ and when $a_{3d}$ is negative. 

In addition to treating the 1d many-body Hamiltonian using the
(FN-)DMC technique, we solve the 1d Schr\"odinger equation 
using the LL equation of state ($g_{1d}>0$) and the 
hard-rod equation of state ($g_{1d}<0$) within the LDA 
(see Sec.~\ref{theoryn}). 
These treatments are expected to be good
when the size of the cloud is much larger than the harmonic oscillator length 
$a_z$, where $a_z =\sqrt{\hbar/m \omega_z}$, that is,
when $a_{3d}$ is large and positive or when $a_{3d}$ is
negative. 

Dotted lines in Fig.~\ref{fig4} show the 1d energy per particle calculated
within the LDA for $g_{1d}^0$ (using the LL equation of
state),
while 
solid lines show the 1d energy per particle calculated
within the LDA
for $g_{1d}$, Eq.~(\protect\ref{eq_g1dren})
(using the LL equation of
state for $g_{1d}>0$, and the hard-rod equation of state for
$g_{1d}<0$).
Remarkably, the LDA energies  
nearly coincide with the 1d many-body DMC energies calculated using 
the unrenormalized coupling constant
(pluses) and the renormalized coupling constant (squares),
respectively.
Finite-size effects play a minor role only for
$a_{3d}\ll a_\rho$. Our calculations thus establish 
that a simple treatment, i.e., a hard-rod equation of state 
treated within the LDA, describes inhomogeneous quasi-1d 
Bose gases with negative coupling constant $g_{1d}$ 
well over a wide range of 3d scattering lengths $a_{3d}$. 

For $a_{3d} \rightarrow 0^-$, 
that is, for large $a_{1d}$,
the hard-rod equation of state treated within the LDA, 
cannot properly describe trapped
quasi-1d Bose gases, which are expected to become unstable
against formation of cluster-like
many-body bound states for
$a_{1d} \approx 1/n_{1d}$.
Thus, Sec.~\ref{stability} investigates 
the regime with negative $a_{3d}$ in more detail
within a many-body framework.

\section{Stability of quasi-one-dimensional Bose gases}
\label{stability}
This section discusses the stability of {\em{inhomogeneous}}
quasi-1d Bose gases with negative $g_{1d}$, that is, with 
$a_{3d}>a_{3d}^c$ and $a_{3d}<0$, against cluster formation.
Section~\ref{energeticsn} shows that the 
(FN-)DMC results 
for the 1d Hamiltonian, Eq. (\ref{eq_h1d}),
are in very good agreement with the 
FN-DMC 
results for the 3d Hamiltonian.
Hence, we carry 
our analysis out within the 1d model Hamiltonian, 
Eq.~(\ref{eq_h1d}); we believe that our final 
conclusions also hold for the 3d Hamiltonian, Eq.~(\ref{eq_h3d}).
For the inhomogeneous 1d Hamiltonian $H_{1d}$, 
Eq.~(\ref{eq_h1d}), the lowest-lying gas-like state is
a highly-excited state (see Sec.~\ref{theoryn}). 
We now
address the question whether this state is stable
quantitatively using the VMC method. 

We solve the 1d many-body Schr\"odinger equation for 
the Hamiltonian $H_{1d}$, Eq.~(\ref{eq_h1d}), by the
VMC method using the trial wave function $\psi_T$ given by 
Eqs.~(\ref{VMC4}) and (\ref{VMC5}).
This many-body wave function  
has the same nodal constraint as a system of $N$ hard-rods of size $a_{1d}$.
However, 
contrary to hard-rods, for 
interparticle distances smaller than $a_{1d}$ 
the amplitude of the wave function increases as $|z|$ decreases. 
This effect arises from the attractive nature of the 1d effective potential and gives rise 
in the many-body framework
to the formation of cluster-like bound states as the average interparticle distance is reduced 
below a certain critical value.

\begin{figure}[tbp]
\vspace*{-1.0in}
\centerline{\epsfxsize=3.25in\epsfbox{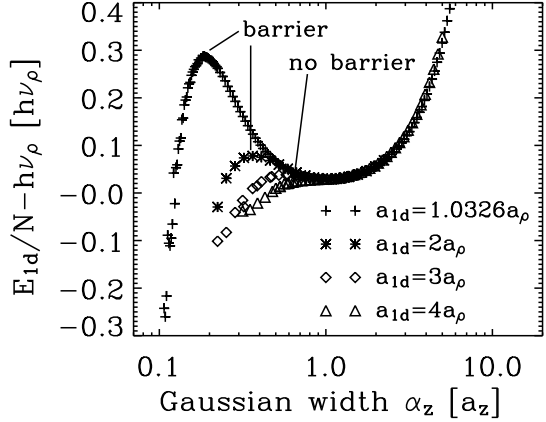}}
\vspace*{-.3in}
\caption{
VMC energy per particle, $E_{1d}/N-\hbar \omega_{\rho}$, 
as a function of the variational
parameter $\alpha_z$ for $N=5$,
$\lambda=0.01$ 
and $a_{1d}/a_{\rho}=1.0326$ (pluses), $2$ (asterisks), $3$ (diamonds)
and $4$ (triangles).
An energy barrier is present for $a_{1d}/a_{\rho}=1.0326$ and 2, but
not for $a_{1d}/a_{\rho}=4$.
}
\label{fig_stab1}
\end{figure}

\begin{figure}[tbp]
\vspace*{-1.0in}
\centerline{\epsfxsize=3.25in\epsfbox{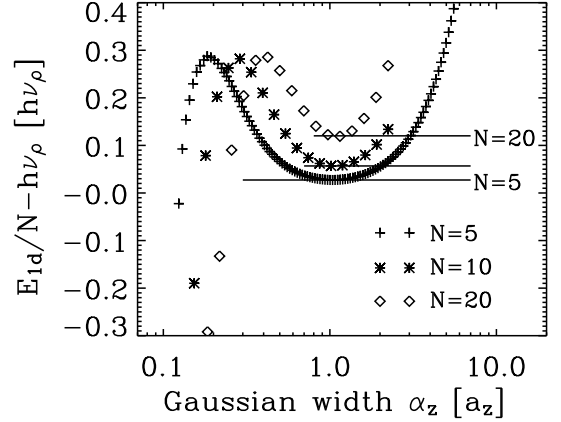}}
\vspace*{-.3in}
\caption{
VMC energy per particle, $E_{1d}/N-\hbar \omega_{\rho}$, 
as a function of the variational
parameter $\alpha_z$ for $a_{1d}/a_{\rho}=1.0326$ (corresponding to
the unitary regime),
$\lambda=0.01$, 
and $N=5$ (pluses), $10$ (asterisks), and $20$ (diamonds).
(The $N=5$ data are also shown in Fig.~\protect\ref{fig_stab1}.)
The height of the energy barrier decreases with increasing $N$.
Horizontal solid lines show the corresponding energies 
for $N=5$, $10$ and $20$ obtained
using the FN-DMC technique, which are in excellent
agreement with the VMC energy obtained for $\alpha_z=\alpha_{z,min}$.
}
\label{fig_stab2}
\end{figure}

Figure~\ref{fig_stab1} 
shows the resulting VMC energy per particle, 
$E_{1d}/N- \hbar \omega_{\rho}$, for $N=5$ and 
$\lambda=0.01$ as a function of the Gaussian width $\alpha_z$ for four different values of $a_{1d}$.
For $a_{1d}/a_{\rho}=1.0326$ and $2$, Fig.~\ref{fig_stab1} 
shows a local minimum at $\alpha_{z,min} \approx a_{z}$. 
The minimum VMC energy nearly coincides with the FN-DMC energy (see 
also Fig.~\ref{fig_stab2}), which 
suggests that our variational 
wave function 
provides a highly accurate description of the quasi-1d many-body system.
The energy barrier at $\alpha_z \approx 0.2 a_{z}$ decreases with increasing $a_{1d}$, and disappears for 
$a_{1d}/a_{\rho}\approx 3$. 
We interpret this vanishing of the energy barrier as an indication of 
instability~\cite{Bohn}. For small $a_{1d}$, the energy barrier separates the 
lowest-lying gas-like state from 
cluster-like bound states. Hence, the gas-like state is stable against 
cluster formation. 
For larger $a_{1d}$, this energy barrier disappears 
and the gas-like state becomes unstable against cluster formation. 

We stress that our stability analysis
should not be confused with that 
carried out for attractive inhomogeneous 3d systems at the level
of mean-field Gross-Pitaevskii theory~\cite{Baym}. 
In fact, a mean-field type analysis of inhomogeneous 1d Bose gases
does not predict stability of gas-like states~\cite{Carr1}. In our analysis,
the emergence of local energy minima in configuration space is
due to the structure of the two-body correlation factor $f_2(z)$ 
entering the VMC trial wave function $\psi_T$, 
Eqs.~(\ref{VMC4}) and (\ref{VMC5}). 
It is a many-body effect that cannot be described within
a mean-field Gross-Pitaevskii framework.

To additionally investigate the dependence
of stability on the number of particles,
Fig.~\ref{fig_stab2} 
shows the VMC energy for $\lambda=0.01$ as a function of the
variational parameter $\alpha_z$
for different values of $N$, $N=5,10$ and $20$.
The scattering length $a_{1d}$ is fixed at the value 
corresponding to the unitary regime, $a_{1d}=1.0326 a_{\rho}$. 
Figure~\ref{fig_stab2} shows that the height of the energy barrier decreases
for increasing $N$.
Figures~\ref{fig_stab1} and \ref{fig_stab2} suggest that the stability of 
1d Bose gases depends on $a_{1d}$ and $N$. 
To extract a functional dependence,
we additionally perform variational 
calculations for larger $N$ and different values of $\lambda$ and $a_{1d}$. 
We find that the onset of instability  
of the lowest-lying 
gas-like state can be described by the following criticality
condition
\begin{equation}
\sqrt{N\lambda}\frac{a_{1d}}{a_\rho}
\simeq 0.78 \;,
\label{stability2}
\end{equation}
or, equivalently, by $\sqrt{N}{a_{1d}}/{a_z}
\simeq 0.78$.
Our 1d many-body calculations thus suggest that
the lowest-lying gas-like state is stable 
if $\sqrt{N\lambda}a_{1d}/a_\rho\lesssim 0.78$,
and that it is unstable if
$\sqrt{N\lambda}a_{1d}/a_\rho\gtrsim 0.78$.
The stability condition, Eq. (\ref{stability2}), implies
that reducing the anisotropy parameter $\lambda$ should allow
stabilization of relatively large quasi-1d Bose gases.

\begin{figure}[tbp]
\vspace*{-1.0in}
\centerline{\epsfxsize=3.25in\epsfbox{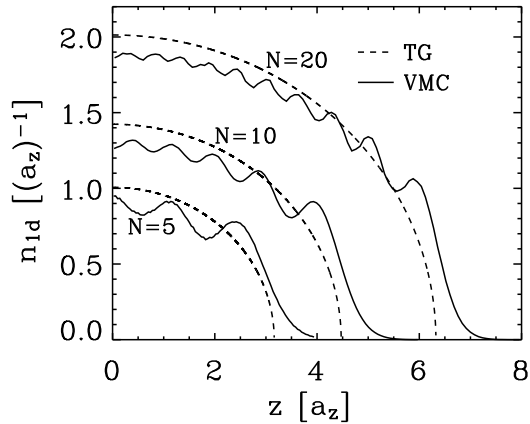}}
\vspace*{-.3in}
\caption{
TG density [Eq.~(\protect\ref{eq_ntg}), dashed lines] as a function of $z$
together with VMC
density (solid lines), obtained by solving the 1d many-body Schr\"odinger equation,
Eq.~(\protect\ref{eq_se1d}), for $N=5$ and $a_{1d}/a_{\rho}=3.6$, 
for $N=10$ and $a_{1d}/a_{\rho}=2.6$, and for $N=20$ and $a_{1d}/a_{\rho}=1.8$. 
The TG density at the center of the trap, $z=0$, deviates from the VMC
density at the center of the trap by less than 10~\%.}
\label{fig_stab3}
\end{figure}

To express the stability condition, Eq.~(\ref{stability2}), in terms 
of the 1d gas parameter $n_{1d} a_{1d}$, where $n_{1d}$ denotes
the linear density at the trap center,
we approximate the density for negative $g_{1d}$ by the 
TG density, Eq.~(\ref{eq_ntg}). Figure~\ref{fig_stab3} 
compares the TG density with that
obtained from the VMC calculations for $N=5, 10$ and 20 and values of 
$a_{1d}/a_{\rho}$ close to the criticality condition, 
Eq.~(\ref{stability2}). 
The density at the center of the trap is described by the TG density to
within 10~\%. Since the TG density at the trap center is given by 
$\sqrt{2N}/(\pi a_z)$ [see Eq.~(\ref{eq_ntg})],
the stability condition, Eq.~(\ref{stability2}), 
expressed in terms of the 1d gas parameter
reads $n_{1d}a_{1d}\lesssim 0.35$.

\section{Conclusions}
\label{conclusions}

This paper presents a thorough study of the properties of 
inhomogeneous, harmonically-confined quasi-1d 
Bose gases as a function of the 3d scattering length $a_{3d}$. 
The behavior of confined Bose gases strongly depends on the ratio of the 
harmonic oscillator length in the tight transverse direction, $a_\rho$, 
to the interaction 
range $r_0$ and to the average interparticle distance $1/n^{1/3}$, where $n$ denotes the 3d central density. 

Quasi-1d bosonic gases have been realized experimentally in highly-elongated 
harmonic traps.
The strength of atom-atom interactions can be varied over a wide range by 
tuning the value of the 3d $s$-wave scattering length $a_{3d}$ 
through application of an 
external magnetic field in the proximity of a Feshbach resonance. 
For $r_0\ll a_\rho$, the 
scattering length $a_{3d}$ determines to a good
approximation the effective 1d scattering length 
$a_{1d}$ and the effective 1d coupling 
constant $g_{1d}$,
which can be, just as the 3d coupling constant,  
tuned to essentially any value including zero and $\pm\infty$.
By exploiting Feshbach resonance techniques, 
one should be able to achieve strongly-correlated 
quasi-1d systems.
The strong coupling regime is achieved for $1/n^{1/3}\gg a_\rho$, it 
includes the TG gas, where a system of interacting bosons behaves as 
if it consisted of 
non-interacting spinless fermions, and the so-called unitary regime, where  
the properties of the gas become independent of the actual value of $a_{3d}$.
In the unitary regime, the gas is dilute, that is, $nr_0^3\ll 1$, 
and at the same time strongly-correlated, that is, $n|a_{3d}|^3\gg 1$.

The present analysis is carried out within various theoretical frameworks. We obtain the 3d
energetics of the lowest-lying gas-like state of the system using a microscopic FN-DMC approach, 
which accounts for all degrees of freedom explicitly. The resulting energetics are then used to 
benchmark our 1d calculations. Full microscopic 1d calculations for contact interactions with 
renormalized coupling constant $g_{1d}$ result in energies that are in excellent agreement with 
the full 3d energies. This agreement implies that a properly chosen many-body 1d Hamiltonian 
describes quasi-1d Bose gases well.

We also consider the LL and the hard-rod equation of state of a 1d system treated within the LDA. 
These approaches provide a good description of the energy of the lowest-lying 
gas-like state for as few as five or ten particles. Finite size effects are to a good approximation 
negligible. Our detailed microscopic studies suggest that these LDA treatments provide a good 
description of quasi-1d Bose gases. In particular, we suggest 
a simple treatment 
of 1d systems with negative $g_{1d}$ using the hard-rod equation of state.

Finally, we address the question of whether the lowest-lying gas-like state of inhomogeneous quasi-1d 
Bose gases is actually stable. We find, utilizing a variational 1d many-body framework, that the 
lowest-lying gas-like state is stable for negative coupling constants, up to a minimum critical value 
of $|g_{1d}|$. Our numerical results suggest that the stability condition can
be expressed as $n_{1d} a_{1d} \simeq0.35$. 
Since our conclusions are derived from variational 1d 
calculations, more thorough microscopic calculations are needed to confirm our findings. We believe,
however, that our findings will hold even in a 3d framework or when three-body
recombination effects are included explicitly.

While our study was performed for inhomogeneous quasi-1d Bose gases, many findings also apply to 
homogeneous quasi-1d Bose gases. 
Furthermore, the Fermi-Bose mapping~\cite{Girardeau,Granger,fermibose}, 
which allows one to map an 
interacting 1d gas of spin-polarized fermions to an interacting 1d gas of spin-polarized bosons, 
suggests that many of the results presented here for quasi-1d Bose gases may directly apply to quasi-1d 
Fermi gases.

{\em Acknowledgments:}
GEA and SG acknowledge support by the Ministero dell'Istruzione,
dell'Universit\`a e della Ricerca (MIUR). DB acknowledges support by
the NSF (grants 0331529 and 0218643) and generous hospitality by the BEC Center at the
University of Trento. BEG acknowledges support by the NSF through
a grant to ITAMP.

\end{document}